\documentclass[a4paper,fleqn]{article}
 \usepackage{amsmath,amsfonts,cite}
\usepackage[nodayofweek]{datetime}

\def\bea{\begin{eqnarray}}
\def\eea{\end{eqnarray}}
\def \pd {\partial}
\newtheorem{theorem}{Theorem}

\begin{document}

\noindent{\bf \Large Differential equations invariant under conditional symmetries} \\

\noindent {\bf   Decio Levi$^*$, Miguel A. Rodr\'iguez$^{\dag}$, Zora Thomova$^+$ }

\noindent{\bf  $^*$ 
  INFN,  Sezione  Roma Tre, Via della Vasca Navale 84, 00146 Roma, Italy, Email: levi@roma.infn.it\\
  $^{\dag}$ Departamento de F\'isica Te\'orica, Universidad Complutense, 28040 Madrid, Spain, Email: rodrigue@fis.ucm.es\\
  $^+$ SUNY Polytechnic Institute,                  
100 Seymour Road,                       
Utica, NY 13502, USA, Email: Zora.Thomova@sunyit.edu                              

}

\begin{center}
{\bf \shortdate
\today}
\end{center}
\begin{abstract}
 Nonlinear PDE's having {\bf given} conditional symmetries are constructed. They are obtained starting from the invariants of the {\it conditional symmetry} generator  and imposing  the extra condition given by  the characteristic of the symmetry. Series of examples starting from the Boussinesq and including non-autonomous Korteweg--de Vries like equations are given to show and clarify the methodology introduced.
\end{abstract}
\section{Introduction}
As Galileo Galilei said in  {\bf Il Saggiatore }(1623) \cite{gg}, our world is described in mathematical formulas and it is up to us to comprehend it. This was the starting point of the scientific revolution which goes on up to nowadays and gave us the present world technology, i.e. cellular phones, lasers, computers, nuclear resonance imaging, etc. .

Our capability of solving complicated physical problems described by mathematical formulas (say equations) is based  on the existence of symmetries, i.e. transformations which leave the equations invariant. Towards the end of the nineteenth century, Sophus Lie introduced the notion of Lie group of symmetries in order to study the solutions of  differential equations. He showed the following main property:  if an equation  is invariant under a one-parameter Lie group of point transformations then we can construct  an invariant solution. This observation unified and extended the available integration techniques as separation of variables or integrating factors. Roughly speaking,  Lie point symmetries  are a local group of transformations which map every solution of the system into another solution of the same system. In other words, it maps the solution set of the equation into itself.

A partial differential equation (PDE) $\mathcal E=0$ is invariant under a symmetry group if the corresponding infinitesimal symmetry generator $\hat X$ is such that 
\bea \label{1.1}
 \qquad \qquad \qquad \mbox{pr} \hat X \mathcal E \Big |_{\mathcal E=0}=0,
 \eea
 where by the symbol {\bf pr} we mean the prolongation of the infinitesimal generator to all  derivatives appearing in the equation $\mathcal E=0$. In particular, if we consider a second order PDE in $\mathbb{R}^2$ of independent variables $x$ and $y$ and dependent variable $u(x,y)$,
 \bea \label{1.1a}
\qquad \qquad  \mathcal E = \mathcal E(x,y,u, u_x, u_y, u_{xx}, u_{xy}, u_{yy})=0,
 \eea
(where the subscripts denote partial derivatives) the infinitesimal generator will be given by
 \bea \label{1.2}
\qquad \qquad  \hat X = \xi(x,y,u)\partial_x+ \eta(x,y,u)\partial_y +\phi(x,y,u)\partial_u,
 \eea
 where $\xi$, $\eta$ and $\phi$ are functions of their arguments to be determined by solving \eqref{1.1}. The prolongation of $\hat X$ is given by
 \bea \label{1.3}
 \mbox{pr}\hat X &=& \hat X + \phi^{(1,x)}(x,y,u,u_x,u_y)\partial_{u_x}+\phi^{(1,y)}(x,y,u,u_x,u_y)\partial_{u_y}+\\ \nonumber &+&\phi^{(2,xx)}(x,y,u,u_x,u_y,u_{xx},u_{xy},u_{yy})\partial_{u_{xx}}+\\ \nonumber &+&\phi^{(2,xy)}(x,y,u,u_x,u_y,u_{xx},u_{xy},u_{yy})\partial_{u_{xy}}+\\ \nonumber &+&\phi^{(2,yy)}(x,y,u,u_x,u_y,u_{xx},u_{xy},u_{yy})\partial_{u_{yy}},
 \eea
 where the functions $\phi^{(1,x)}$,  $\phi^{(1,y)}$ and $\phi^{(2,xx)}$,  $\phi^{(2,xy)}$, $\phi^{(2,yy)}$ are algorithmically  derived in terms of $\xi$, $\eta$ and $\phi$. See,  for example, the references \cite{olver,bk,ba,bca,h,ref3,stephani} for this construction.
 
A function $\mathcal I$ is an invariant of a symmetry if it is such that
\bea \label{1.4}
\qquad \qquad \mbox{pr}\hat X \mathcal I=0.
\eea
Eq. \eqref{1.4} is a first order PDE which can be solved on the characteristic and provide  the set of invariants $\mathcal I_j$, $j=0,1, \cdots$ depending on $x$, $y$, $u$ and its partial derivatives up to the second order.  Then a PDE invariant with respect to the infinitesimal generator \eqref{1.2} can be written as
\bea \label{1.5}
\qquad \qquad \mathcal E=\mathcal E(\{\mathcal I_j\})=0, \qquad j=0,1, \cdots .
\eea

Lie method is a well established technique to search for exact solutions of differential or difference equations of any type, integrable or non--integrable, linear or nonlinear. However, many equations may have no symmetries and there is no simple algorithm to prove the existence of symmetries other than looking for them. Moreover, the obtained solutions do not always fulfill the conditions imposed by the physical requests (boundary conditions, asymptotic behavior, etc.). So one  looks for extension or modification of the construction which could overcome some of these problems. One looks for more {\sl symmetries},  
\begin{itemize}
\item not always expressed in local form in terms of the dependent variable of the differential equations,
\item not satisfying all the properties  of a Lie group but just providing solutions.
\end{itemize}
In the first class are the potential symmetries introduced by Bluman et. al.  \cite{bpot,bkr},  the nonlocal symmetries by Vinogradov et. al. \cite{13,25,hr,21,22,mr,cf,nlr} while in the second one are the conditional symmetries \cite{bluman,levi4,f,nc,fz}.

In this paper we  will be interested in showing that one can construct equations having given conditional symmetries.

In Section 2 we will provide the theory behind  the construction of the conditional symmetries clarifying in this way the difference between symmetries and conditional symmetries. Then in Section 3 we will verify the proposed construction in the case of the Boussinesq equation \eqref{3.1} and, in correspondence with its conditional symmetries, construct new conditionally invariant equations.   Section 4 is devoted to  the summary of the result, some concluding remarks and prospects of future works. 
\section{What is a conditional symmetry?}
Conditional symmetries were introduced by  Bluman and Cole with the name {\it non-classical method} \cite{bluman}  
by adding an auxiliary first-order equation to \eqref{1.1a}, build up in terms of the coefficients of the infinitesimal generator $\hat X$, namely
\bea \label{2.1}
\mathcal C = \mathcal C(x, y, u, u_x, u_y)=  \xi(x,y,u) u_x+ \eta(x,y,u) u_y -\phi(x,y,u)=0,
\eea
the infinitesimal symmetry generator \eqref{1.2} written in characteristic form \cite{olver} set equal to zero. Equation \eqref{2.1} is as yet unspecified and it will be determined together with the vector
field $\hat X$, as it involves the same functions $\xi, \, \eta$ and $\phi$. I.e. we look for the simultaneous symmetry group of the overdetermined system of  equations \eqref{1.1a} and \eqref{2.1}. It is easy to prove that  \eqref{2.1} is invariant under the first prolongation of \eqref{1.2} 
\bea \label{2.1a}
\qquad \qquad \mbox{pr} \hat X \mathcal C =-(\xi_{u} u_x + \eta_{u} u_y -\phi_u) \,\mathcal C,
\eea
 without imposing any conditions on the functions $\xi, \, \eta$ and $\phi$. Consequently, we  need just to apply the following invariance condition
\bea \label{2.2}
\qquad \qquad  \mbox{pr} \hat X \mathcal E \Big |_{\substack{\mathcal E=0 \\[1pt] \mathcal C=0}}=0.
 \eea
Eq. \eqref{2.2} gives  nonlinear determining equations for $\xi$, $\eta$ and $\phi$ which provide at the same time the classical and non--classical symmetries. In fact, as noted in \cite{clarkson}, since all solutions of the classical determining equations necessarily satisfy the nonclassical determining equations \eqref{2.2}, the solution set may be larger in the nonclassical case. As $\mathcal C=0$ appears in \eqref{2.2} as a condition imposed on the determining equations one has  called the resulting symmetries {\it conditional symmetries}. 

There are several works devoted to using the non-classical method to construct solutions of PDEs that are different from the ones obtained by classical method using the Lie point symmetries.  Among them, let us cite as an example,  \cite{cm,by,piv,rkfs,QuJS,agb,gs,aabd,hn}. In the case of integrable equations let us mention the works of Sergyeyev \cite{18,19} where he considered  the classification  of all ($1 + 1$)-dimensional evolution systems that admit a generalized (Lie--B\"acklund) vector field as a generalized conditional symmetry.

In this paper we want to look at the conditional symmetries from a different perspective. Given an infinitesimal group generator characterized by a vector field $\hat X$ for specific values of the functions $\xi, \, \eta$ and $\phi$,  we want to construct equations $\mathcal E=0$ which have this symmetry as a {\it conditional symmetry} and not as a Lie point symmetry. Taking into account that an equation invariant under a given symmetry is written in terms of its invariants  \eqref{1.5}, a second order PDE invariant under a conditional symmetry will be given by
\bea \label{2.3}
\qquad \qquad \mathcal E(\{\mathcal I_j \}) \Big |_{\{\mathcal C=0\}}=0, \qquad j=0,1,\cdots.
\eea
The constraint  $\{\mathcal C=0\}$ in \eqref{2.3} is to be interpreted as the differential equation \eqref{2.1} and all of its differential  consequences (see Section 3 for the details presented in the explicit examples).

The  condition $\mathcal C=0$ and its differential consequences  must not be used  everywhere on the invariant equation  to get  \eqref{2.2} as, if we would do so, 
the global substitution of the condition and its consequences would turn the invariant PDE into an ODE in one of the independent variables  with parametric dependence on the other.


\section{A series of examples including the Boussinesq equation.}
The Boussinesq equation  
\bea \label{3.1}
\qquad \qquad u_{yy} + u u_{xx} + (u_x)^2 + u_{xxxx}=0,
\eea
was introduced in 1871 by Boussinesq to describe  the propagation of long waves in shallow water \cite{bou1,bou2} and it is of considerable physical and mathematical interest.  It also arises in several other physical applications including one-dimensional nonlinear
lattice waves\cite{za,toda}, vibrations in a nonlinear string\cite{zakharov}, and ion
sound waves in a plasma\cite{scott}.

If $\eta$ in \eqref{1.2} is different from zero the resulting determining equations for conditional symmetries do not fix it and we can always put it equal to one. The same phenomena happens when $\eta=0$ and we have $\xi \ne 0$. In this case we can put $\xi =1$.

The conditional symmetries of the Boussinesq equation for $\eta \neq 0$ were obtained in \cite{levi4}, and  in \cite{ck} by non group techniques. The case $\eta=0$ has been considered later and can be found in  \cite{clarkson}. Moreover it is worthwhile to notice that the condition is the same if we consider $\hat X$ or $f(x,y,u) \hat X$, however the invariants in the two cases are different. So, for any $\hat X$ we can consider $f(x,y,u) \, \mathcal C$ as a condition.

In \cite{levi4,clarkson} we find the following generators of the conditional symmetries for \eqref{3.1}:
\bea \label{3.2}
\hat X_1&=&\partial_y+y \partial_x-2y \partial_u
\\ \label{3.3}
\hat X_2&=&\partial_y-\frac{x}{y} \partial_x+\Big (\frac 2 y u + \frac 6 {y^3} x^2 \Big ) \partial_u
\\ \label{3.4}
\hat X_3&=&\partial_y+\Big (-\frac{x}{y} +y^4 \Big )\partial_x+\Big (\frac 2 y u + \frac 6 {y^3} x^2 -2 y^2 x -4 y^7\Big ) \partial_u
\\ \label{3.5}
\hat X_4&=&\partial_y+\Big (\frac{x}{2y} +y \Big )\partial_x- \frac 1 y ( u + 2x +4 y^2  ) \partial_u
\\ \label{3.6}
\hat X_5&=&\partial_y+\frac{1}{2} \frac {\dot \wp} \wp (x+\beta_2 W) \partial_x-\Big [\frac {\dot \wp}{\wp} u + 3 \dot \wp x^2 + \frac {\beta_2}{2}  \Big ( \frac 1 \wp + 12 \dot \wp W \Big )  \\ \nonumber &+&\frac{\beta_2^2} 2  W \Big ( \frac 1 \wp + 6 \dot \wp W \Big ) \Big ] \partial_u,  \quad W(y) = \int_0^y \frac{\wp (s)}{[\dot \wp (s)]^2} ds, 
\\ \label{3.7}
\hat X_6&=&\partial_x+ \Big [\frac 2 {x+c_0} u + \frac{48}{(x+c_0)^3} \Big ] \partial_u,
\\ \label{3.8}
\hat X_7&=&\partial_x+ \Big [-2xQ+c_1 Q + c_2 Q \int_0^y\frac{ds}{[Q(s)]^2} \Big ]\partial_u,
\eea
where $\wp$ is a special case of the Weierstrass elliptic function $\wp (y,g_2,g_3)$ \cite{as} with $g_2=0$ satisfying the differential equations $\dot \wp^2=4\wp^3-g_3$, $Q=\wp(y+c_3,0,g_3)$ and $\beta_2$, $g_3$ and $c_i,\;i=0,\ldots, 3$  are arbitrary constants.

The generators $\hat X_1, \cdots, \hat X_5$ were obtained assuming $\eta=1$, thus are defined in (\ref{3.2}-\ref{3.6}) up to an arbitrary function $\eta(x,y,u)$ while $\hat X_6$ and $\hat X_7$ were obtained assuming $\eta=0$ and $\xi =1$, thus are defined in \eqref{3.7} and \eqref{3.8} up to an arbitrary function $\xi(x,y,u)$.

\subsection{Conditional invariant equations associated to $\hat X_1$}
For the infinitesimal generator $\hat X_1$ and its prolongation up to fourth order we obtain the following invariants:
\bea \nonumber
\mathcal I_0&=& -2x+y^2, \quad \mathcal I_1= 2  x + u, \quad \mathcal I_2= u_x, \quad \mathcal I_3=2y+yu_x+u_y,\quad \mathcal I_4=  u_{xx},\\ \nonumber
 \mathcal I_5&=& y u_{xx}+u_{xy}, \quad \mathcal I_6=u_{yy} +2 y u_{xy}+2 (y^2 -x ) u_{xx}, \quad \mathcal I_{7}=  u_{xxx}\\ \label{3.1.1}
& \ldots&, \; \mathcal I_{11}=  u_{xxxx}.
\eea
The condition is given by $\mathcal I_3=0$ i.e. $\mathcal C=2y+yu_x+u_y=0$ and we can construct the Boussinesq equation in terms of the invariants \eqref{3.1.1}. It is:
\bea \label{3.1.2}
\qquad \mathcal I_1 \mathcal I_4 +  \mathcal I_6 +\mathcal I_{11}+\mathcal I_2^2  \Big |_{\mathcal C_x=0}=0.
\eea  
As $ {\mathcal C}_x=u_{xy}+y u_{xx}$ we have:
\bea \label{3.1.2a}
&&u_{yy}+2y(u_{xy}+yu_{xx})+u u_{xx}+u_{xxxx}+(u_x)^2 \Big |_{u_{xy}+y u_{xx}=0}=\\\nonumber &&=u_{yy} + u u_{xx}  + u_{xxxx}+ (u_x)^2=0.
\eea
So, the Boussinesq equation \eqref{3.1} has the conditional symmetry given by $\hat X_1$.

Now, we want to construct an autonomous equation which have the conditional symmetry given by $\hat X_1$. Let us consider a different subset  of the invariants \eqref{3.1.1} 
\bea \nonumber
&& \mathcal I_6 +\mathcal I_1 \mathcal I_4 \Big |_{\mathcal C_x=0 }=0, \, \mbox{i.e.} \;
\Big [ u_{yy} +2y(u_{xy}+yu_{xx})+u u_{xx} \Big ] \Big |_{u_{xy}+y u_{xx}=0}=0.
\eea  
So we we have:
\bea \label{3.1.4}
\qquad \qquad u u_{xx} + u_{yy}=0,
\eea
a nonlinear Laplace equation which is a truncation of the Boussinesq equation.

To verify if effectively \eqref{3.1.4} has $\hat X_1$ as a conditional symmetry we  compute its Lie point symmetries. They   are
\bea \label{3.1.5}
\hat Z_1&=&\pd_x\\ \nonumber
\hat Z_2&=& \pd_y \\ \nonumber
\hat Z_3&=& x \pd_x+y\pd_y.
\eea
So   \eqref{3.1.4} does  have $\hat X_1$ as a conditional symmetry. 

A KdV like non-autonomous equation which has the conditional symmetry given by $\hat X_1$ can be found by considering a different subset  of the invariants \eqref{3.1.1}, i.e. 
\bea \label{3.1.6}
&& \mathcal I_7 + (\mathcal I_1  +\mathcal I_0) \mathcal I_2   \Big |_{\mathcal C=0 }=0, \; \mbox{i.e.} \;
\Big ( u_{xxx}+u u_x + y^2  u_x \Big ) \Big |_{y u_x=- u_y - 2 y}=0.
\eea  
We get:
\bea \label{3.1.7}
\qquad \qquad u_y  = \frac 1 y ( u_{xxx} + u u_x ) -2 y.
\eea

To verify if effectively \eqref{3.1.7} has $\hat X_1$ as a conditional symmetry we  compute its Lie point symmetries. They   are
\bea \label{3.1.8}
\hat Z_4&=&\pd_x\\ \nonumber
\hat Z_3&=&\ln y  \pd_x - \pd_u \\ \nonumber
\hat Z_2&=& [y^2 \ln y +\frac{x}{3}-\frac{y^2}{6} ] \pd_x + y \ln y \pd_y - [ 2 y^2 \ln y + \frac{2y^2}{3}-\frac{2u}{3}]\pd_u  \\ \nonumber
\hat Z_1&=&y^2 \pd_x +y \pd_y-2y^2 \pd_u= y \hat X_1.
\eea
The invariants of $\hat{Z_1}$ are:
\bea \label{3.1.8a}
I_0 &=& 2  x + u, \quad I_1= -2x+y^2, \quad  I_2= u_x, \quad  I_3=yu_y+2xu_x+4x, \\ \nonumber  
I_4 &=&  u_{xx}, \quad I_5 = 2 x u_{xx}+yu_{xy}, \\ \nonumber
I_6 &=& y^2 u_{yy} +4 x y u_{xy}+4 x^2 u_{xx}+2 x u_x+4x, \quad  I_{7}=  u_{xxx},  \ldots
\eea
and \eqref{3.1.7} is given by $I_0I_2-I_3+I_7-2I_1=0$. The equation \eqref{3.1.7} is not conditionally invariant under the field $\hat{X_1}$.

\subsection{Conditional invariant equations associated to $\hat X_2$}
For the infinitesimal generator $\hat X_2$ and its prolongation up to derivatives of fourth order we obtain the following invariants:
\bea\label{3.2.1}
&&\mathcal I_0= xy, \quad \mathcal I_1= x^2 u + \frac{x^4}{y^2}, \quad \mathcal I_2= x^3 u_x + 2 \frac{x^4}{y^2}, \\ \nonumber &&\mathcal I_3=\frac{x^2}{y} (y u_y - x u_x -2 u -6 \frac{x^2}{y^2}), \quad \mathcal I_4=  x^4 u_{xx} + 2 \frac{x^4}{y^2},\\ \nonumber 
&& \mathcal I_5= \frac{x^3} y \Big (yu_{xy}-xu_{xx} -3 u_x -12 \frac x {y^2}\Big ), \\ \nonumber  &&\mathcal I_6= \frac{x^2} y \Big (y u_{yy}  -2 x u_{xy} + \frac{x^2} y u_{xx} -4 u_y +6 \frac x y  u_x +6 \frac u y + 42 \frac{x^2}{y^3} \Big ),  \\ \nonumber  &&\mathcal I_{7}=  x^5 u_{xxx},  \ldots,\; \mathcal I_{11}=  x^6 u_{xxxx}. 
\eea
The condition is given by $\mathcal I_3=0$ i.e. $\mathcal C=y u_y - x u_x -2 u -6 \frac{x^2}{y^2}=0$ and we can construct a nonlinear evolution PDE in terms of the invariants \eqref{3.2.1}. It is:
\bea \label{3.2.2}
\qquad  \mathcal I_1 \mathcal I_4 +\mathcal I_{11} + \mathcal I_2^2   ={x}^{6} \left [ u_{{yy}}+uu_{{xx}}+u_{{xxxx}}+(u_{{x}})^{2}
 \right ],
\eea 
when $\mathcal C=0$ and all its differential consequences.  

A KdV like non-autonomous equation which may have the conditional symmetry given by $\hat X_2$ can be found by considering a different subset of  the invariants \eqref{3.2.1}, i.e. 
\bea \label{3.2.3}
&& \mathcal I_7 +\mathcal I_1   \mathcal I_2  \Big |_{\mathcal C=0 }=0, 
\eea  
that is
\bea \label{3.2.4}
\qquad \qquad u_y  + \frac y x ( u_{xxx} + u u_x ) - 4 \frac {x^2} {y^3}=0.
\eea
Lie point symmetries of \eqref{3.2.4} are
\bea \label{3.2.5}
\hat Z_1&=&y \pd_y+\frac{1}{2}x\pd_x-u\pd_u \\ \nonumber
\hat Z_2&=&\frac{1}{y^5} \Big[\pd_y-\frac{x}{y}\pd_x+\Big(\frac{2}{y}u+\frac{6}{y^3}x^2 \Big)\pd_u \Big]\\ 
\hat Z_3&=&y^7\pd_y+2y^6x\pd_x-4y^4 (y^2u-3x^2  )\pd_u \nonumber
\eea
We observe that $\hat Z_3=\frac{1}{y^5} \hat X_2$.
The invariants of $\hat Z_3$ are:
\bea \label{3.2.6}
&&I_0=xy,\quad I_1=x^2(u+\frac{x^2}{y^2}),\quad I_2=x^3(u_x+\frac{2x}{y^2}),\\ \nonumber && I_3=x^7(u_y+\frac{x}{y}u_x-\frac{2}{y}u-\frac{6x^2}{y^3}), \quad I_4=x^4(u_{xx}+\frac{2}{y^2}),\, \ldots \\ \nonumber && I_7=x^5 u_{xxx},
\eea
and \eqref{3.2.4} is given by $I_3+I_0I_7+I_0 I_1 I_2 =0$. Thus the equation \eqref{3.2.4} is invariant under the vector field $\hat Z_3$ of \eqref{3.2.5} and not conditionally invariant under the vector field $\hat X_2$.
\subsection{Conditional invariant equations associated to $\hat X_3$}
For the infinitesimal generator $\hat X_3$ and its prolongation up to derivatives of fourth order we obtain the following invariants:
\bea\label{3.3.1}
&&\mathcal I_0= xy-\frac 1 6 y^6, \quad \mathcal I_1= \frac u {y^2}  + \frac{x^2}{y^4} - \frac 1 3 x y + \frac{13}{18} y^6, \quad \mathcal I_2= \frac{u_x}{y^3}  + 2 \frac{x}{y^5}, \\ \nonumber &&\mathcal I_3=\frac{1}{y^3} [4 y^8 + y^5 u_x +2 y^3 x +y u_y - x u_x -2 u -6 \frac{x^2}{y^2}], \quad \mathcal I_4=  \frac{u_{xx}}{y^4}  + 2 \frac{1}{y^6},
 \\ \nonumber  && \mathcal I_{7}= \frac{u_{3x}} {y^5},  \ldots,\; \mathcal I_{11}=  \frac{u_{xxxx}}{y^6}, \ldots \; .
\eea

The condition is given by $\mathcal I_3=0$ i.e. $\mathcal C=4 y^8 + y^5 u_x +2 y^3 x +y u_y - x u_x -2 u -6 \frac{x^2}{y^2}=0$ and we can construct a nonlinear evolution PDE in terms of the invariants \eqref{3.3.1}, provided the condition is satisfied. It is:
\bea \label{3.3.2}
 &&\mathcal I_2 (\mathcal I_2 -9) +\mathcal I_{11} +\mathcal I_4 ( \mathcal I_1-\frac 5 3 \mathcal I_0)-36 =\\ \nonumber && \qquad \qquad =\frac 1{y^6} \left[ u_{{yy}}+uu_{{xx}}+u_{{xxxx}}+(u_{{x}})^{2}
 \right],
\eea  
when $\mathcal C=0$ and all its differential consequences. 

Then, the Boussinesq equation \eqref{3.1} has the conditional symmetry given by the vector field $\hat X_3$.

A KdV like non-autonomous equation which may have the conditional symmetry given by $\hat X_3$ can be found by considering a different subset of the invariants \eqref{3.3.1}, i.e. 
\bea \label{3.3.3}
&&\qquad \qquad \qquad \Big (\frac{637}{75} \mathcal I_0+\mathcal I_1 \Big )\mathcal I_2+\mathcal I_7-\frac{8}{25} \mathcal I_1  \Big |_{\mathcal C=0 }= 0
\eea
We get:
\bea \label{3.3.4}
xu_{{y}}+y[u_{{xxx}}+uu_{{x}}]={\frac {49}{75}}\,{y}^{4} \,u-y^4{
\Big (\frac {229}{25}}\,x
-{\frac {52}{75}}\,{y}^{5} \Big )u_{{x}}-{\frac {53}{3}}\,{x}^{2}{y}^{2}+4\,{\frac {{x}^{3}}{{y}^{3}}}.
\eea
The equation \eqref{3.3.4} has no point symmetries. Then  \eqref{3.3.4}  is a conditionally invariant KdV-like equation.
\subsection{Conditional invariant equations associated to $\hat X_6$} \label{x6}
For the infinitesimal generator $\hat X_6$ with $c_0=0$ and its prolongation up to second order we obtain the following invariants:
\bea\label{3.4.1}
&&\mathcal I_0= y, \qquad  \mathcal I_1= \frac {1}{x^4} \Big (12+ x^2 u \Big ), \qquad  \mathcal I_2=  \frac{u_y}{x^2}, \\ \nonumber &&\mathcal I_3=\frac{1}{x^5} \Big ( x^3 u_x - 2 x^2  u  - 48  \Big ), \qquad \mathcal I_4=   \frac{u_{yy}}{x^2}, \cdots , \\ \nonumber &&\mathcal I_{10}=\frac 1 {x^7} \Big ( -1440 -24 x^2 u +18 x^3 u_x + x^5 u_{xxx} - 6 x^4 u_{xx} \Big ).
\eea
The condition is given by $\mathcal I_3=0$ i.e. $\mathcal C=u_x -  \frac{2 u} x   - \frac{48}{x^3}=0$ and we can construct a nonlinear evolution PDE in terms of the invariants \eqref{3.4.1}. It is:
\bea \label{3.4.2}
6 \mathcal I_1^2 +   \mathcal I_4   =\frac 1{x^2} \left[ u_{{yy}}+uu_{{xx}}+u_{{xxxx}}+(u_{{x}})^{2}
 \right],
\eea  
when $\mathcal C=0$ and all its differential consequences.

Then, the Boussinesq equation \eqref{3.1} has the conditional symmetry given by $\hat X_6$.

A KdV like non-autonomous equation which may have the conditional symmetry given by $\hat X_6$ can be found by considering a different subset of  the invariants of \eqref{3.4.1}, i.e. 
\bea \label{3.4.3}
&& \mathcal I_3^2 +\mathcal I_2     + \mathcal I_{10}  \Big |_{\substack{\mathcal C=0 \\[1pt] {{\mathcal C}_x}=0}}=0.\eea  
Eq. \eqref{3.4.3} reads:
\bea \label{3.4.5}
&& u_{y}+u_{{xxx}} - 4 \, {\frac{u u_x}{x^3}}+8\,{\frac {{u}^{2}}{{x}^{4}}}+192\,{\frac {u}{{x}^{6}}}- \frac{288}{{x}^{5}} =0
\eea
Equation \eqref{3.4.5} has only Lie point symmetry $\hat{Z_1}=\pd_y$.
\bea \label{3.4.6}
\hat Z_1&=&\pd_y 
\eea
So, \eqref{3.4.5} is a conditionally invariant equation.
\subsection{Conditional invariant equation associated to \newline \indent $  \hat Y=\partial_y + \frac x {2y} \partial_x - \frac 1 y \partial_u$}
The generator $\hat Y=\partial_y + \frac x {2y} \partial_x - \frac 1 y \partial_u$ was introduced by Momoniat \cite{mo} to describe the nonclassical (conditional) symmetries of the Frank-Kamenetskii partial differential equation 
\bea \label{3.5.1a}
\frac{\partial u}{\partial y}= \frac 1 x \, \frac {\partial}{\partial x} \left (  x \frac {\partial u}{\partial x} \right ) + e^u,
\eea
modeling a thermal explosion in a cylindrical vessel. In \cite{mo} the obtained symmetry for \eqref{3.5.1a} was shown to correspond to a classical symmetry.

For the infinitesimal generator $\hat Y$ and its prolongation up to third order we obtain the following invariants:
\bea \nonumber
\mathcal I_0&=&\frac  y {x^2}, \quad \mathcal I_1= 2 \ln x + u, \quad \mathcal I_2= x u_x, \quad \mathcal I_3=\frac 1 {2y}(2+xu_x+2yu_y),\\ \label{3.5.1}
\mathcal I_4&=& x^2 u_{xx}, \; \ldots, \quad \mathcal I_6 = u_{yy} +\frac x y u_{xy} + \frac 1 4 \frac {x^2}{y^2} u_{xx} - \frac 1 4 \frac x {y^2} u_x - \frac 1 {y^2}, \\ \nonumber 
\mathcal I_7&=& x^3 u_{xxx}.
\eea
The condition is given by $\mathcal I_3=0$ i.e. $2+xu_x+2yu_y=0$ and we can construct, apart from the Frank-Kamenetskii partial differential equation \eqref{3.5.1a},  a nonlinear KdV like  evolution PDE in terms of the invariants \eqref{3.5.1}. It is:
\bea \label{3.5.2}
\qquad \mathcal I_0 e^{\mathcal I_1} +  \mathcal I_0 \mathcal I_7 + 1 + \frac 1 2 \mathcal I_2  \Big |_{\mathcal I_3=0}=0, \quad  \mbox{i.e.} \quad u_y=   x u_{xxx} + e^u
\eea
i.e. a nonlinear dispersive non-autonomous KdV like equation.
Lie point symmetries of \eqref{3.5.2} are
\bea \label{3.5.3}
\hat Z_1&=&\pd_y\\ \nonumber
\hat Z_2&=& y \pd_y+\frac{1}{2}x\pd_x-\pd_u
\eea
We observe that $\hat Z_2=y \hat Y$. 
The invariants of $\hat Z_2$ are:
\bea \label{3.5.4}
I_0={\frac {y}{{x}^{2}}},\; I_1=2\,\ln   x  +u,\; I_2=x \,u_{{x}},\;I_3={x}^
{2}\,u_{{y}},\ldots,\; I_7={x}^
{3}\,u_{{xxx}},
\eea
and \eqref{3.5.2} is given by $I_3-I_7-\exp(I_1) =0$. Thus the equation \eqref{3.5.2} is invariant under $\hat Z_2$  and not  conditionally invariant under  $\hat Y$. 


\section{Conclusions}
In this article we presented a construction of nonlinear PDE's having {\bf given} conditional symmetries. They are obtained starting from the invariants of the symmetry  and imposing  the extra condition given by equating to zero the characteristic and its differential consequences. Starting from the conditional symmetries of the Boussinesq equation we re-construct the Boussinesq equation itself as well as other  nonlinear equations  
(\ref{3.1.4}, \ref{3.1.7}, \ref{3.2.4}, \ref{3.3.4}, \ref{3.4.5}, \ref{3.5.2}). Equations (\ref{3.1.7}, \ref{3.2.4}, \ref{3.3.4}, \ref{3.4.5}, \ref{3.5.2}) are non-autonomous KdV--like equations and it is well known that the KdV equation has no conditional symmetries \cite{ck}. However, not all obtained equations are conditionally invariant even if we constructed them in such a way. The obtained equations can still have the generator $\hat X$ as a point symmetry due to the arbitrary multiplicative factor $\eta(x,y,u)$ or $\xi(x,y,u)$ under which the condition is defined. This is what happens in cases of the KdV-like equations  (\ref{3.1.7}, \ref{3.2.4}, \ref{3.5.2}). 

An important point not touched in this work but on which we are presently working is understanding a priori when we can construct a conditionally invariant equation;  why most of the KdV like equation we have constructed are not conditionally invariant?  Moreover 
work is also  in progress on solving by symmetry reduction the obtained conditionally invariant KdV like equations and on the construction of  conditional symmetry preserving discretizations of the Boussinesq equation.

\paragraph{Acknowledgments.}   DL has been supported by  INFN IS-CSN4 {\it Mathematical Methods of Nonlinear
Physics}. DL thanks ZT and the SUNY Polytechnic Institute for their warm hospitality at Utica when this work was started. DL thanks the Departamento de F\'isica T\'eorica  of the Complutense University in Madrid for its hospitality. MAR was supported by the Spanish MINECO under project FIS 2015-63966-P. D. Nedza, summer student of ZT, contributed to the verification of some of the computations.


\end{document}